\documentclass[journal]{IEEEtran}

\usepackage{color}
\usepackage[inline]{enumitem}
\usepackage{scrextend}
\usepackage[noadjust]{cite}
\usepackage{fancyvrb}
\usepackage{lipsum}
\usepackage{wrapfig}
\usepackage{amssymb}
\usepackage{authblk}
\usepackage{scalerel}
\usepackage{mathcomp}

\usepackage{adjustbox}

\usepackage{graphicx}

\graphicspath{{./pdf/}{./jpeg/}{./png/}}
\DeclareGraphicsExtensions{.pdf,.jpeg,.png}
\usepackage{amsmath}
\usepackage{algorithm2e}
\usepackage{url}
\usepackage{booktabs}

\begin{document}
%
\title{\emph{MedShare:} Medical Resource Sharing\\ among Autonomous Healthcare Providers}
%
%
%




\author{Yilong~Yang,~\IEEEmembership{Student Member,~IEEE,}
~Xiaoshan~Li,
Nafees~Qamar,
Wei~Ke,
and~Zhiming~Liu
\thanks{Yilong~Yang and Xiaoshan~Li are with the Faculty of Science and Technology, University of Macau, Macau. E-mail: yylonly@gmail.com, xsl@umac.mo.}
\thanks{Nafees~Qamar is with Department of Computer Science, Biomedical and Health Informatics program, State University of New York, USA. Email: nafees.qamar@oswego.edu.}
\thanks{Wei~Ke is with Macau Polytechnic Institute, Macau.}
\thanks{Zhiming~Liu is with School of Computer and Information Science, Southwest University, Chongqing, China. Email: zhimingliu88@swu.edu.cn.}}


%
%

\markboth{Journal of \LaTeX\ Class Files,~Vol.~14, No.~8, August~2015}%
{Shell \MakeLowercase{\textit{et al.}}: Bare Demo of IEEEtran.cls for IEEE Journals}
%



\maketitle

\begin{abstract}
Legacy Electronic Health Records (EHRs) systems were not developed with the level of connectivity expected from them nowadays. Therefore, interoperability weakness inherent in the legacy systems can result in poor patient care and waste of financial resources. Large hospitals are less likely to share their data with external hospitals due to economic and political reasons. Motivated by these facts, we aim to provide a set of software implementation guidelines, i.e., \emph{MedShare} to deal with interoperability issues among disconnected healthcare systems. The proposed integrated architecture includes: 1) a data extractor to fetch legacy medical data from a hemodialysis center, 2) converting it to a common data model, 3) indexing patient information using the HashMap technique, and 4) a set of services and tools that can be installed as a coherent environment on top of stand-alone EHRs systems. Our work enabled three cooperating but autonomous hospitals to mutually exchange medical data and helped them develop a common reference architecture. It lets stakeholders retain control over their patient data, winning the trust and confidence much needed towards a successful deployment of \emph{MedShare}. Security concerns were effectively addressed that also included patient consent in the data exchange process. Thereby, the implemented toolset offered a collaborative environment to share EHRs by the healthcare providers.
\end{abstract}


%

\section{Introduction}
\IEEEPARstart{L}{egacy} EHRs systems have been mostly designed and implemented to meet the internal clinical needs of healthcare providers, which have become obsolete and no longer meet the external needs of the patients and local governments. Consequently, this impedes the way to an improved patientcare in a networked healthcare setting, also resulting in increased cost and clinical negligence. The future health information systems aimed at the integration, interoperability, innovation, and intelligence \cite{Greenes2016}\cite{Murua2018} for sharing the resource. Exchanging medical information has seamlessly paved the way to introducing medical standards \cite{Garde2007}\cite{ Bakken2000}\cite{adams2017analysis} providing with a unified approach to medical vocabulary and exchange of information, but none of them has come of age to be used smoothly. For example, a study~\cite{doi:10.1093/jamia/ocv103} finds weak evidence of the `meaningful use program (MU)' initiated by the 2009 Health Information Technology for Economic and Clinical Health (HITECH) Act on EHRs uptake due to data interoperability challenges. The study \cite{TURAN201457} presents the top ten technical issues in healthcare, which include privacy, quantity, security, and the implementation of electronic medical records. Moreover, the political and economical issues and healthcare providers of contingent factors \cite{NGUYEN2014779} should take into an account in the development of medical information sharing.

Large medicalcare providers seem reluctant to share their patient cum customers data with other healthcare providers \cite{miller2014health}. They exchange patient information internally and are less likely to cooperate outside their network \cite{RePEc:eee:jhecon:v:33:y:2014:i:c:p:28-42}. In such a scenario, the design and development of an interoperable health information exchange system is a non-trivial task. This is not only because of complex workflows involving data acquisition, storing, communication, and manipulation, but also lacking in a coordinated effort to connect autonomous healthcare providers.

Albeit, in an ideal scenario healthcare networks are expected to: (a) to support direct data exchange, (b) query-based exchange of patient-related information in an emergency situation, medication history, radiology reports and records of a diseased person hospitalized for emergency care, and (c) personalized patient data management by patients themselves like online banking. Architecting and implementing such an interoperable system, meeting the aforementioned requirements, needs a comprehensive and multifaceted approach to catering both technical non-technical issues. 

Motivated by this, the current research is focused on connecting disintegrated healthcare providers in Macau SAR that include three major hospitals named Hospital Conde S. Janu\'{a}rio (HC), Kiang Wu Hospital (KW) and Macau University of Science and Technology Hospital (UH). However, the theme of our work has wider implications and scope to build health information exchange systems that confront the same challenges. The autonomous EHRs systems under consideration were neither developed using special instructions or standards at the time of their birth, nor the concerned authorities were ready to update their legacy systems. Because the three hemodialysis centers had their fully functional and independent electronic health records in place. Among the three collaborating hospitals in this research, two are private healthcare providers while the third one is public.

\begin{figure*}[!htb]
\centering
\includegraphics[width=0.8\textwidth]{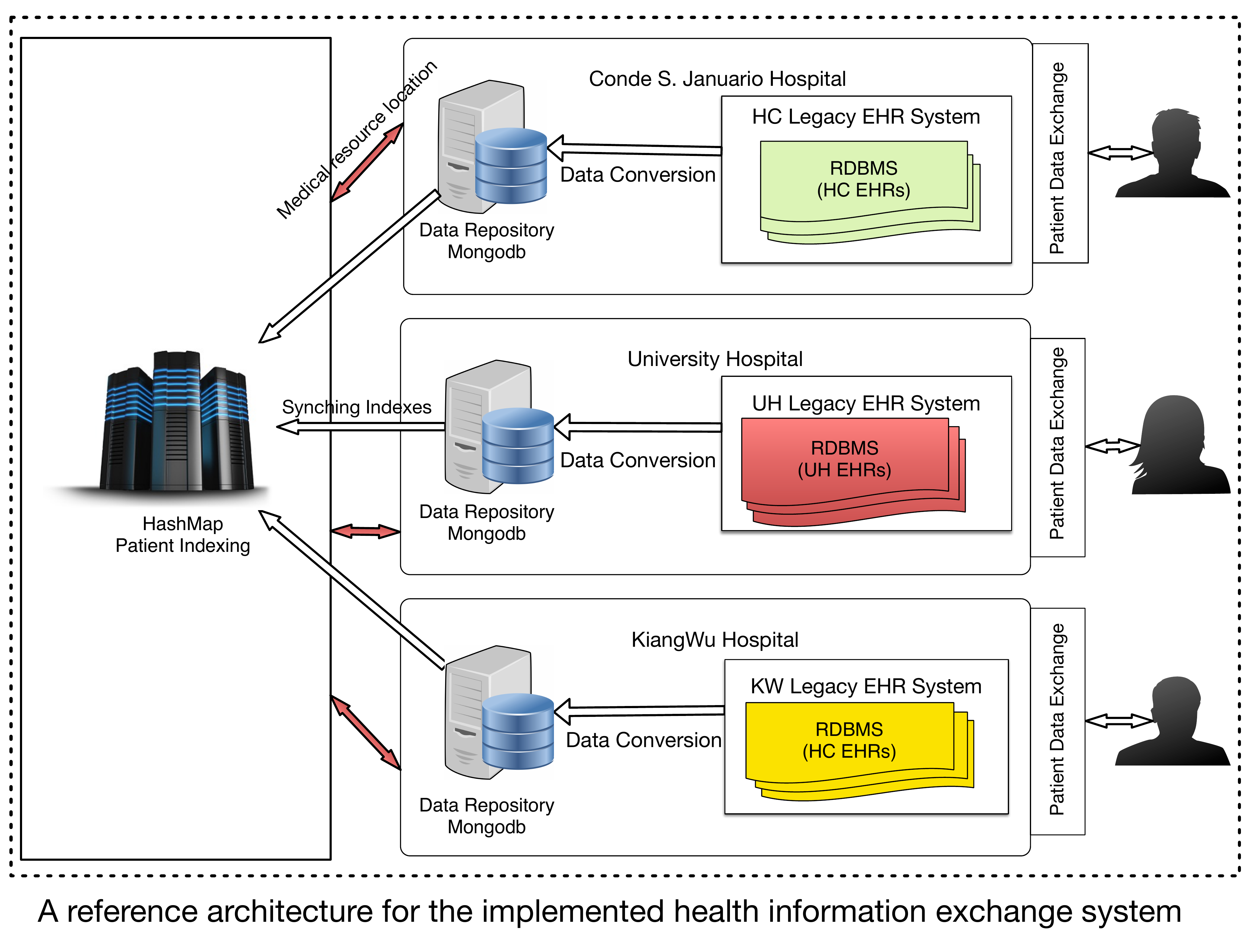}\\
\caption{A High Level View of the \emph{MedShare} Architecture}
\label{ucdarch}
\end{figure*}

In the given healthcare setting, distributed information sharing is mandatory for effective patient care and monitoring where patients often opt for switching a healthcare provider due to numerous reasons. \emph{MedShare} is a simple yet robust EHRs system to allow for exchanging medical resources for an improved patient care between isolated hemodialysis centers in the given scenario. The types of data shared in \emph{MedShare} includes lab reports, radiology images, transcription reports and medication histories. \emph{MedShare} works in three steps: 1) it uses a data extractor to extract legacy data of a patient located at a hemodialysis centers, 2) it converts the data to a unified data format agreed upon by all the stakeholders and medical providers belonging to three hospitals, 3) the platform indexes the patient information using the HashMap technique. Our approach integrates a set of services and tools that can be installed as a coherent environment on top of standalone EHRs. As discussed in~\cite{HYPPONEN20141}, Operational Data Model (ODM) lacks explicit support for modern exchange mechanisms, and our authentication mechanism is based on RESTful web services and our previous work also employs the same techniques to exchange medical information~\cite{qamar2016querying}. The \emph{MedShare} EHRs sharing system, as depicted in Fig. \ref{ucdarch}, allows to handle situations such as follows:

\noindent \textbf{Example}: A doctor can request all the hemodialysis records of a patient. The EHR sharing system returns a date-wise list of all the hemodialysis records of a queried patient. Furthermore, the doctor can access the detail of the EHRs on a specific date. E.g., Sep 30, 2015. \emph{MedShare} allows an administrator to track the potential leaks in the system. For example, when the tracker needs to know the accessing information about the EHR with ID 0221, the tracking system shows all the relevant results. Hence, \emph{MedShare} facilitates with distributed patient care but also allows to share tasks of the hemodialysis EHRs, if required, among the Macau hospitals without compromising patient privacy. We identify the data exchange scenarios, capture the intent behind it and identify collaborating entities. These and other system goals are achieved by system components such as authentication, EHR query, synchronization, and audit.

\vspace{.1cm}
\noindent\textbf{Contributions:} This paper offers three substantive contributions. 1) We set up a reference architecture for a diverse set of healthcare providers to connect and exchange medical information of their patients. 2) The implemented approach is reusable. The source code of the system is uploaded on GitHub\footnote{\url{https://github.com/yylonly/medshare}}, which is freely available to download, and 3) the implementation sets forth technological guidelines for designing and implementing health information exchange system.

For brevity, the remainder of the paper is organized as follows: Section~\ref{sec:Relatedwork} reviews the related literature. Section~\ref{sec:SharingPattern} presents data exchange scenarios from the hemodialysis centers in Macau. Section~\ref{sec:architecure} proposes \emph{MedShare}, a medical data resource sharing architecture. Section~\ref{sec:Prototype} shows system prototyping and demonstration. Section~\ref{sec:Conclusion} concludes this paper and outlines our future work.

\section{A Literature Review}
\label{sec:Relatedwork}

Legacy EHR systems were not developed with a certain level of interoperability in mind. Therefore, dis-connectivity inherent in these systems can result in their inability to exchange medical resources. Contrarily, numerous benefits can be achieved by connecting legacy EHRs systems. The authors~\cite{FEATHERSTONE201245} demonstrate a suggestive evidence that a shared electronic health record can support more integrated care. More evidence comes through quantitative analyses of the actual contribution of shared EHR systems and is discussed in a large case study conducted in Austria \cite{RINNER201644}. But, large-scale adoption of such systems is impractical without addressing the privacy and security concerns \cite{REZAEIBAGHA201625}. On the other hand, larger hospital systems generally exchange electronic patient information internally, not with other external hospitals \cite{miller2014health}. It also reveals that larger hospital systems tend to create `information silos', which is a data system that is incapable of reciprocal operations with other hospital systems. The reason is if larger hospitals allow outflow of data they are more likely to loose patients. In such a given situation, the adaptability of open standards for interoperable hospital systems is still far from practice. This situation necessitates the need to engage health informatics researchers and users for a better interconnection among different hospitals. Another study shows that inter-organizational data exchange is one of the most important information system challenges~\cite{HYPPONEN20141}, where it reports on the user experiences with different regional health information exchange systems in Finland.

A recent work \cite{SINACI2013784} combines metadata registries and semantic web technologies to uniquely reference, query and process a Common Data Element (CDE) to enable the syntactic and semantic interoperability. However, this research is limited to the interoperability of medical vocabulary. The survey consists of 43 techniques addressing interoperability issues. Another study~\cite{doi:10.1136/amiajnl-2012-000855} provides interesting findings that 40.7\% (n=1465) of the predefined headings applied in the multi-professional EHR system was shared by two or more professional groups and only 1.7\% (n=62) of the predefined headings were shared by all eight groups. The study \cite{SINACI2013784} creates the Portal of Medical Data Models\footnote{\url{https://medical-data-models.org}} to foster sharing of medical data models. This is achieved by a web front-end that enables users to search, view, download and discuss data models. Some other related work can be found in ~\cite{huang2012hierarchical}\cite{sun2011}\cite{Garde2007}\cite{Sheth2014}\cite{Kotz2015}\cite{Bakken2000}.

Numerous factors, e.g., scalability, heterogeneity, resource management, transparency, openness, performance analysis and synchronization contribute to the development of a dependable EHR system, nonetheless, security may be considered at the core of system properties. Medical resource sharing, if it is between cross-organization or cross-domain, a study considers cross-domain authentication and fine-grained access control~\cite{5196665}. This study discusses an on-demand revocation if any of the two cooperating organizations are unwilling to share data anymore. This may be seen as the flexibility to the notion of security and privacy concerns in a networked healthcare setting. Another approach~\cite{Reicher2016} uses direct messaging, a secure e-mail-like protocol employed to allow the exchange of encrypted health information online. The paper~\cite{doi:10.1093/jamia/ocv038} provides a tool supports a privacy-preserving linkage of electronic health records (EHRs) data across multiple sites in a large metropolitan area in the United States (Chicago, IL). Another research~\cite{Kwon2015} discusses the possibility of attacks on healthcare systems.

Two closely related work, the first is the eMOLST project~\cite{laszewski2011emolst}, which officially supported by the New York State Department of Health (NYS DOH) providers, handles data interoperability through: a) authenticating access to a shared medical resource by applying Single Sign-On (SSO) technique, b) it employs a patient identity source system to assign a unique identifier to a patient and requires extra work to maintain a set of attributes associated with the patient. In contrast, our system computes the hash code of the patient identifies card number, uniquely representing each patient in the EHR sharing system. eMOLST requires the new system portal to be deployed to access the EHRs, while our system is designed to work with EHR legacy system. Our patient indexing component that lets hospitals keep the data by themselves. eMOLST needs to push the data away to a centralized repository. The second work~\cite{fragidis2016integrated} proposed a semi-distributed architecture NEHR to offer EHR sharing in Greece. Every sharing request requires the authentication of the patient in NEHR. However, our MedShare architecture provides two-way authentication to not only need the authentication of the patient but also require the authentication by the data providers. Moreover, our locator service uses the de-identified HashMap to locate the resource, which reduces the risk of privacy breach.

For privacy requirements, a survey~\cite{Sheth2014} across North America, Asia, and Europe shows that data sharing and data breaches are the biggest concerns for the users. Moreover, there is a comparison~\cite{Liu2013} on the effectiveness of the methods used for anonymizing quasi-identifiers to avoid sensitive information, and this quantitative analysis of de-identification shows that de-identifying data provides no guarantee of anonymity. The study~\cite{Angiuli2015} elaborates on the de-identified patient data, and ~\cite{Qamar2013} argues explicit validation of healthcare security policies. \cite{Hydari2015} relates electronic health records and patient safety. Some other works~\cite{DBLP:journals/titb/KilicDE10}\cite{Lee2014} discuss Peer-to-Peer (P2P) networks to support medical data sharing. A research~\cite{7457335} studies the data exchange between patients and healthcare facilities. It further investigates the real-time data synchronization issues. Our work also addresses synchronization challenges, but no real-time data synchronization is needed in our case.

\section{An Overview of the Hemodialysis Centers in Macau}
\label{sec:SharingPattern}
The hemodialysis centers in Macau serve a large number of population, but they are disconnected to share medical records of their patients. A patient, generally tends to see a doctor in a hospital of her choice, is prescribed a hemodialysis treatment plan at specified dates. If a patient suddenly decides to change her hemodialysis center, the exchange of patient information between hemodialysis centers becomes a bottleneck for the smooth delivery of medical services.

The Macau citizens have confidence in public health systems, and they prefer to see a doctor in HC. Consequently, the initial diagnosis records and treatment plans are produced and stored in HC. Nonetheless, a patient may opt to go to another hospital, say UH, to take treatments due to unavailability of resources and the place where they live. The hemodialysis centers have no sharing platform in place. Therefore, it results in carrying paper-based by the patient medical data along with any other electronic data copies on CDs. It is noteworthy that patient privacy is well preserved with respect to the security of the EHR system in HC for a non-closure data agreement exists between KW and UH.

\begin{figure}[!htb]
\begin{center}
\includegraphics[width=1\columnwidth]{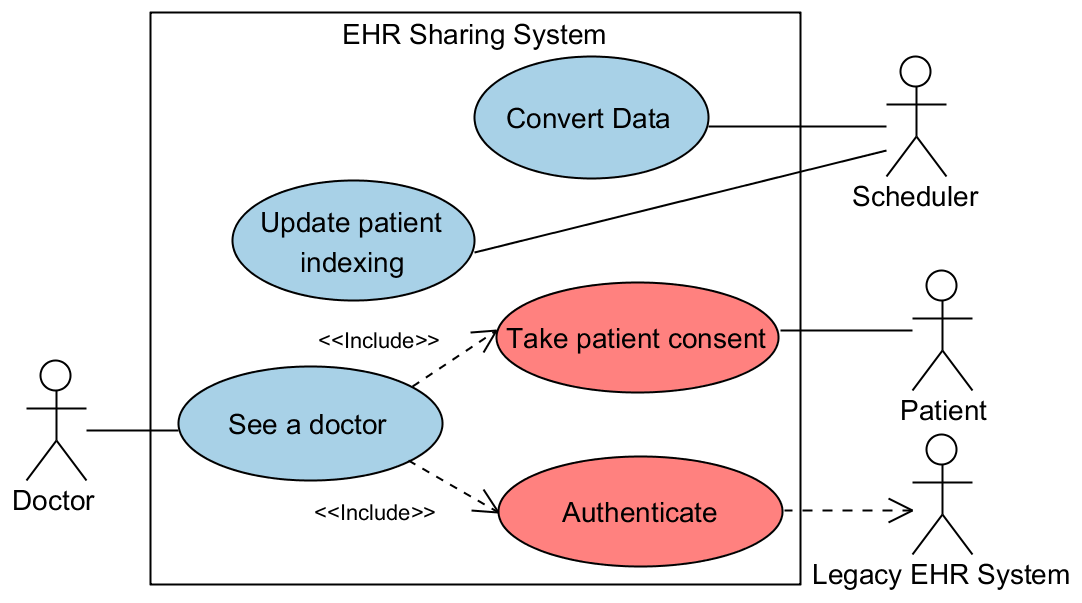}\\
\caption{Use Case Diagram of the networked EHR System}
\label{usecasedia}
\end{center}
\end{figure}

The data-sharing problem leads to developing a hemodialysis network that should address the following functional and non-functional requirements. We use the Unified Modeling Language\footnote{\url{http://www.omg.org/spec/UML/}} (UML) to give a flavor of the EHR system requirements in Fig. \ref{usecasedia}. Some of the main functional requirements are listed below:

\begin{itemize}
\item[--] The use case of seeing a doctor describes the procedure that a patient visits the doctor in a hospital,
and the doctor requests for the related shared EHRs of the patient from other hospitals, if any.
\item[--] A doctor is authenticated and authorized to access a local medical record.
\item[--] A doctor may access medical records placed at another hospital through the same authentication service in her working hospital.
\item[--] The patient provides her consent, and authorizes the doctor to access her medical records. This guarantees that in a EHR sharing session, a patient authorization is recorded.
\item[--] The scheduler updates the local patient data in a unified format and updates it at the indexing server. These shared records should be regularly synchronized but not required to be updated in real-time. Note that a hemodialysis patient usually takes her next treatment after a specificied time.
\end{itemize}

\subsection{The Workflow for Resource Sharing}
The high-level EHR sharing workflow is presented as follows: a patient sees a doctor in an arbitrary hospital $H_1$ among HC, KW and UH.
The EHRs are generated and stored in the respective legacy EHR systems of $H_1$.
A scheduler regularly triggers the synchronization of the extracted shared EHRs from the legacy EHR system
and updates the corresponding indexes in the patient indexing server.
Only then, the patient can see a doctor in another hospital $H_2$ with the access to the shared EHRs.
At the time of requesting old EHRs of the patient, the doctor must be authorized by both the current hospital $H_2$ and the patient.

\begin{figure}[htb]
\begin{center}
\includegraphics[width=0.8\columnwidth]{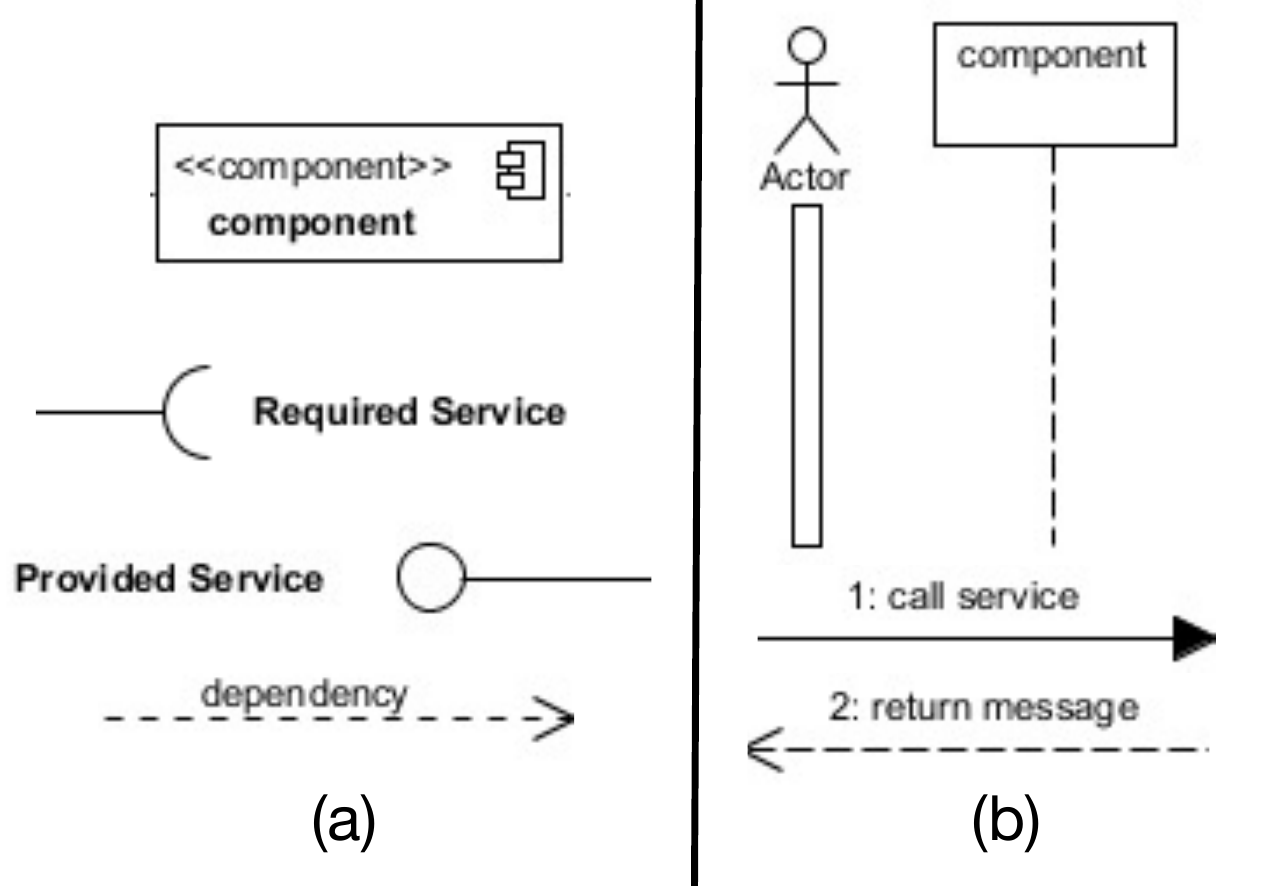}
\caption{The part (a) of the Figure~\ref{umlnotation} shows the self-explanatory UML component diagram and names its elements. Part (b) shows the elements of the sequence diagram presenting the notion of actor and calling functions.}\label{umlnotation}
\end{center}
\end{figure}

\begin{figure*}[!htb]
\begin{center}
\includegraphics[width=0.95\textwidth]{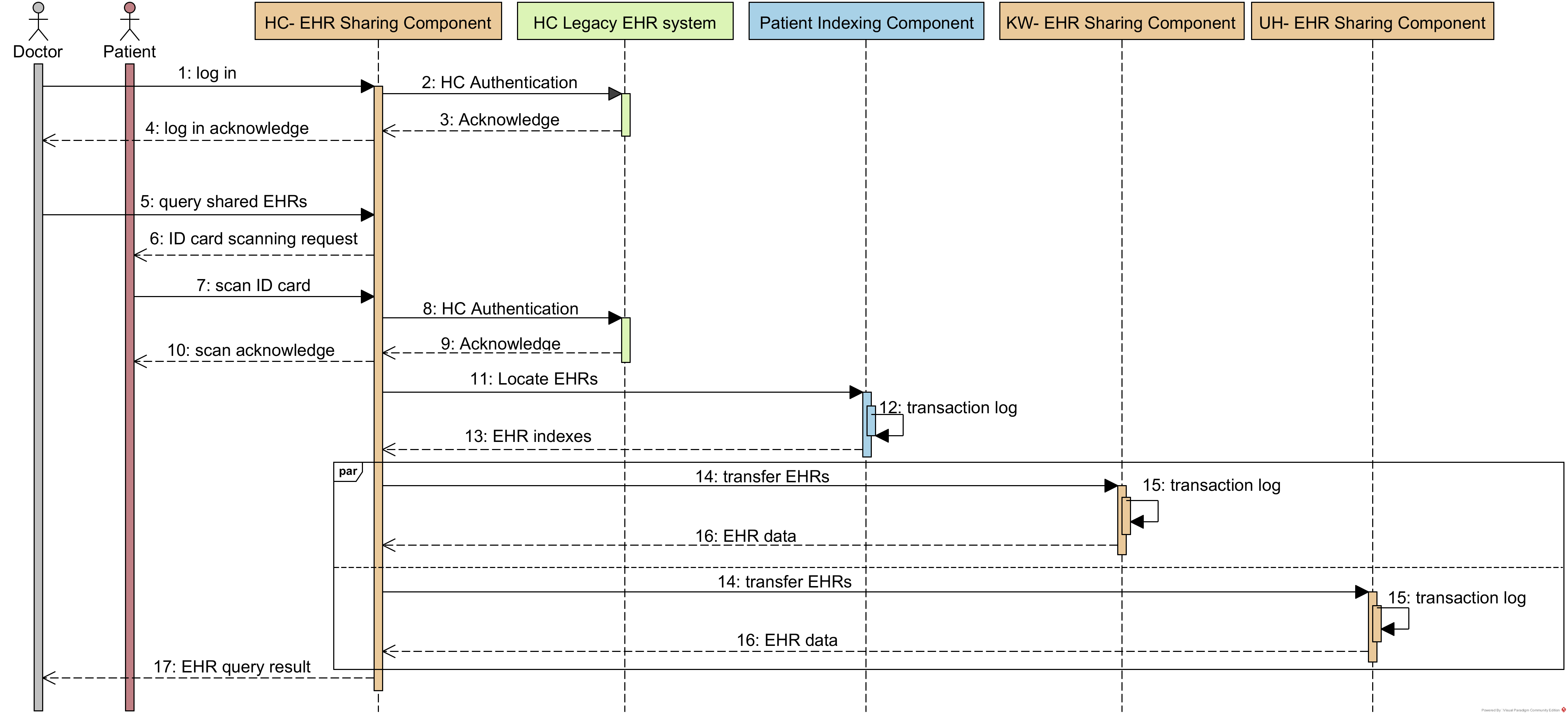}\\
\caption{System Usage Scenarios -- Seeing a Doctor}
\label{seqdia}
\end{center}
\end{figure*}

To understand the graphical notation used in the paper, non-familiar readers are refered to UML specification. However, for brevity, we provide the names and functions of the used notation in Fig.~\ref{umlnotation}. Fig.~\ref{seqdia} presents the detailed system usage scenario of the communication taking place between actors and the EHR sharing system. We use the standard sequence diagram from UML that allows to graphically depict how your system could potentially be interacted with. Considering the proposed architecture, Fig.~\ref{seqdia} describes the detailed activities beyond the system architecture. A doctor of HC is authorized through the service in her working hospital (Please refer to \textit{Steps 1, 2, 3, 4} in the diagram), and then the doctor runs the EHR queries on a patient data (\textit{Step 5}). The patient then authorizes this request by scanning her ID card (\textit{Step 6, 7, 8, 9, 10}). This two-way authentication by the patient and the hospital satisfies the required privacy requirements. The request can then be sent to the data center (\textit{Step 11}).
If the patient data is distributed over multiple locations (e.g. KW and UH), the relevant indexes are retrieved by the query (\textit{Step 13}). Afterwards, the transmission requests will be sent out to those hospitals (\textit{Step 14}).
Once the data transfer (\textit{Step 16}) is completed in EHR sharing client of HC, the requested EHRs are displayed to the doctor (\textit{Step 17}).
Moreover, the transactions are recorded in the log database for post-event analysis (\textit{Step 12, 15}). This is important in case of a privacy and security breach.
If the patient has EHRs in more than one hospital, the operations (\textit{Step 14, 15, 16}) will be run in parallel for each of remote hospitals.

\subsection{Data Format Inconsistencies}
Since the studied legacy EHRs systems were autonomously designed and implemented, a number of database inconsistencies appeared at the time of implementation. The terminologies used to represent the EHR data were not based on any standard or common data format, which needed to be resolved first. TABLE \ref{table1} provides an example of the database entries from the three hospitals, though representing the same meanings, but with different names. The right most column presents the unified format agreed upon by the concerned authorities.

\begin{table}[!htb]
\centering
\caption{An example of unified data format}
\begin{adjustbox}{max width=\columnwidth}
\begin{tabular}{lcccr}
\toprule
Attribute Name & HC Format & KW Format & UH Format & Unified Format \\
\midrule
Patient identity & \verb|card_id| & \verb|identitiy_id| & \verb|id| & \verb|patient_id| \\
EHR identity & \verb|record_id| & \verb|id_ehr| & \verb|eid| & \verb|ehr_id| \\
Patient identity & \verb|p_name| & \verb|patient_n| & \verb|pname| & \verb|patient_name| \\
Name of the doctor & \verb|d_name| & \verb|dotctor_n| & \verb|dname| & \verb|doctor_name| \\
\bottomrule
\end{tabular}
\end{adjustbox}
\label{table1}
\end{table}

The unified EHR data format can significantly reduce the number of data inconsistencies between different EHR formats.
Otherwise, each hospital requires a targeted data conversion for each corresponding hospital.
In our unified EHR sharing scenario, each hospital only requires to conform to a single negotiated data format.
However, EHRs sharing (independent of unified format) requires bidirectional data conversion between two autonomous health care providers and the number of conversions can be calculated by the formula $n(n-1)$ if there are $n$ hospitals. 
However, only $n$ number of conversions are needed in a unified EHR sharing.
Although, we currently have only three hospitals in the Macau EHRs sharing case study, the network
may grow well in the near future and other health providers and research institutes may take part in the data sharing process. In the future, the unified data format will ease the merger of a new healthcare provider into the \emph{MedShare}. Note that the unified data sharing format and the negotiation process was directly held by the administration of the hospitals.
The HL7 format~\cite{Bakken2000}, OpenEHR~\cite{Garde2007} standards, and other semantic models of EHRs \cite{doi:10.1093/jamia/ocv008}\cite{LEGAZGARCIA2016175} were not under consideration during the negotiation process. Our work in this research project was only confined to fill the technological gaps.

\section{Interoperable Architecture for Sharing Medical Resources}
\label{sec:architecure}
This section introduces the architectural aspects of the health information exchange system and elaborates on the technical details encountered in the development of the system. Our experience with developing a large system reveals that interoperability is not  only the issue to enable two autonomous systems to exchange systems, but other non-technical factors also play a vital role. In this regard, one of the challenges lies in mediating the situation when autonomous health care providers are not interested to share the data of their patient cum customers and show a complete lack of interest in transfering the data to their competitors. After presenting the architectural details first, we will present a simple yet robust solution to this problem.

\subsection{MedShare Architecture}
We employ the component diagram based on standardized UML notation of Fig. \ref{umlnotation} (a) to present MedShare Architecture. The architecture has two views: 1) External view: this represents the foundational block of resource sharing approach that allows for linking legacy EHR systems into a collaborative sharing of their data. 2) Internal view: this describes the design for the core components of the MedShare. 



\begin{figure}[!htb]
\begin{center}
\includegraphics[width=\columnwidth]{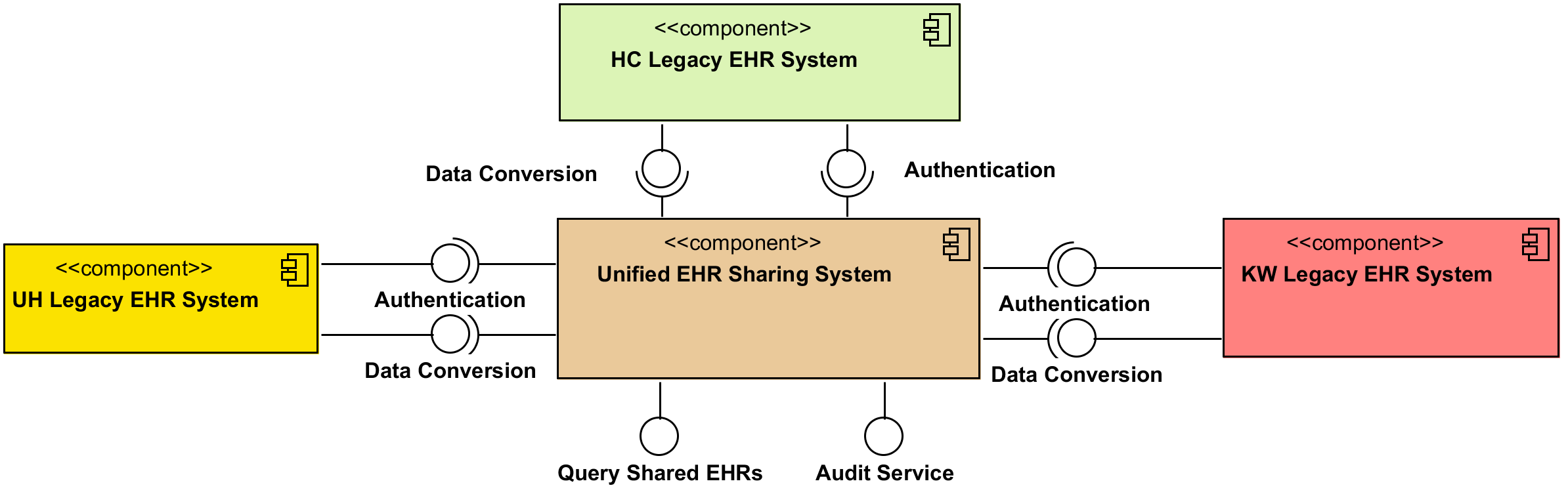}\\
\caption{The External View of Distributed EHR Sharing Architecture}
\label{external}
\end{center}
\end{figure}

\vspace{.3cm}
\noindent
\textbf{External View}: The Figure~\ref{external} illustrates the external view of our system. Legacy EHR systems provide the services of \emph{data conversion} to convert shared EHRs from a legacy system to a distributed EHR system. However, using the \emph{authentication} service the doctor and the patient are authorized. By using both services from the local legacy systems, the unified EHR sharing system provides two services: 1) It allows to run a query on the \emph{MedShare}. 2) The audit service handles the privacy requirements of the system and post-breach data analysis, which is not detailed in this paper.

\begin{figure}[!htb]
\begin{center}
\includegraphics[width=\columnwidth]{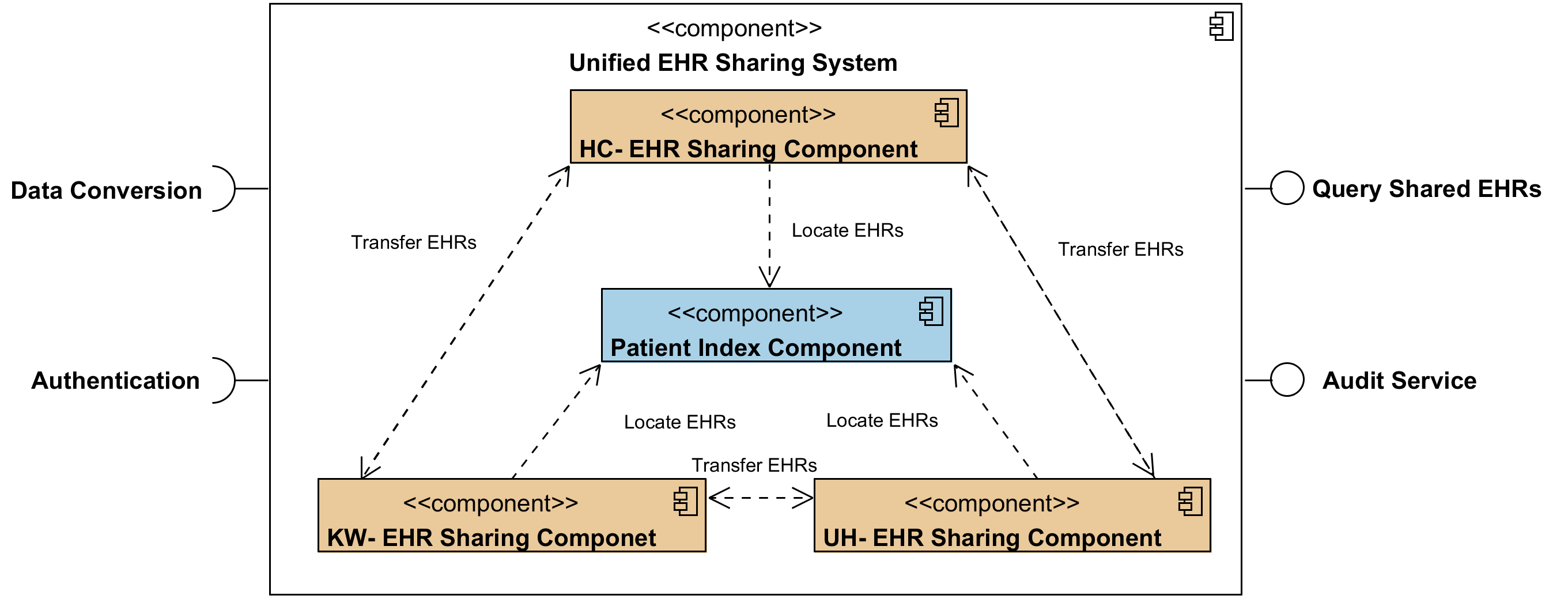}\\
\caption{The Internal View of Distributed EHR Sharing Architecture}
\label{internal}
\end{center}
\end{figure}

\noindent
\textbf{Internal View}:
The internal view of the unified EHR sharing system in Fig.~\ref{internal} shows how the sub-systems collaborate to provide the required medical data querying mechanism from the different hospitals. The subsystems use the services provided by the index system in the data center to locate EHRs, then using the service of \textit{transfer EHRs} in each subsystem to transfer all requested EHRs. To understand what is a service in the system, readers are directed to the implementation section of the paper.

\noindent
\textbf{Patient Indexing}:
The patient indexing component stores all the references of the shared EHRs to facilitate data queries from the participating hospitals, i.e. requesters and providers. A requester poses a data location query to the patient index component without a direct connection with a peer hospital. This is represented as $\dashrightarrow$ with a \emph{locate EHRs} label and shows the dependency between components in Fig.~\ref{internal}. The label \emph{transfer EHRs} provides access to the real data.
The indexing component stores only the unique reference for each shared EHR, without any physical data relocation taking place from the original source. This approach offers two main advantages: 1) huge data synchronization burden is alleviated and 2) cyber-security attacks and other threats from the internal users are minimized.

The \textit{HashMap} technique is employed for patient indexing that includes a relationship between a patient and the EHR with the location. However, we leave it to the healthcare providers to decide about the segments of data to be indexed. Obviously, only the references are not enough. We need to store in the indexing server some attributes of a shared EHR that are not privacy-sensitive, as tags, along with the reference to the EHRs. The indexing server is then able to respond to queries based on these tags. Typically, the tags should include the source location, the encoded patient number, the date and time and the type of the EHRs.
On the principle of facilitating queries while complying privacy policies, it also analyzes which set of tags is to be opened to the indexing server may be pre-negotiated between stakeholders.

There are two main reasons not to use central storage for patient data: 1) a hospital must push all the shared data into the data center before EHR sharing if the data center stores all data and 2) the local data should frequently be synchronized with the indexing server. That will lead to a huge synchronization burden to the data center because of enormous size of data. For example, imagine the CT scan examination report that may contain more than 1GB of data. 

\subsection{Data Query and Output Structure}

As mentioned above, a data query includes two steps: 1) locating an EHR, and 2) the data transfer procedure. An EHR is located by a query, followed by the output. Hereunder, we illustrate this by using an example, which further will be detailed in the implementation section. Below, we provide the query attributes that inlcudes patient identity, choosing the range of dates, EHR type and which hospitals to query.

 \vspace{.1cm}
\noindent
\textbf{Input Parameters for Locating:}

\begin{table}[!htb]
\centering
\normalsize
\begin{tabular}{|c|c|c|c|}
\hline
Hashed Patient ID & Date Range & EHR Type & Hospitals \\
\hline
\end{tabular}
\label{LocatingInput}
\end{table}

\vspace{-0.3cm}
\noindent\textbf{Output:}

\begin{table}[!htb]
\centering
\normalsize
\begin{tabular}{|c|c|c|c|}
\hline
EHR ID & EHR Type & Date & Location \\
\hline
\end{tabular}
\label{LocatingOutput}
\end{table}
Note that the retrieved \textit{ID} in above is used to access a particular EHR resource through \emph{Transfer EHR service} as shown in the Fig.~\ref{internal}.

\vspace{.1cm}
\noindent
\textbf{Input Parameters for Transferring:}

\begin{table}[!htb]
\centering
\normalsize
\begin{tabular}{|c|c|}
\hline
EHR ID & EHR Type \\
\hline
\end{tabular}
\label{TransferringInput}
\end{table}
The desired output is shown in a simplified way as follows. The output shown below is integrated into the graphical user interface of our toolset.

\vspace{.1cm}
\noindent
\textbf{Output:}
\vspace{-.3cm}
\begin{table}[!htb]
\centering
\normalsize
\begin{tabular}{|c|}
\hline
EHR Data \\
\hline
\end{tabular}
\label{TransferringOutput}
\end{table}

\subsection{Ensuring Patient Privacy against Cyber Attacks}

Our proposed technique introduces a two-way authentication process protecting the patient data from cyber-security attacks. A doctor logins into the system, upon which the patient is requested to scan her identity card. This two-way authentication is enforced to take patient consent and protect critical medical resources from outside attacks. In a worst scenario, if the patient indexing server is compromised the hashed patient identities are highly likely to remain protected. The authentication process for doctors is implemented using role-based access control. Note that access to medical data records by a doctor requires laws and regulations to restrain her from any type secondary usage of the data. Above all, all the operations in the resource sharing system are logged to be able to investigate data breaches and perform auditing services.

\section{System Prototyping and Demonstration}
\label{sec:Prototype}
We have put forth a technique that allows for sharing medical resource. In order to realize \emph{MedShare}, we implement our approach in four layers.

\subsection{MedShare Implementation Stack}
\begin{figure}[!htb]
\begin{center}
\includegraphics[width=1\columnwidth]{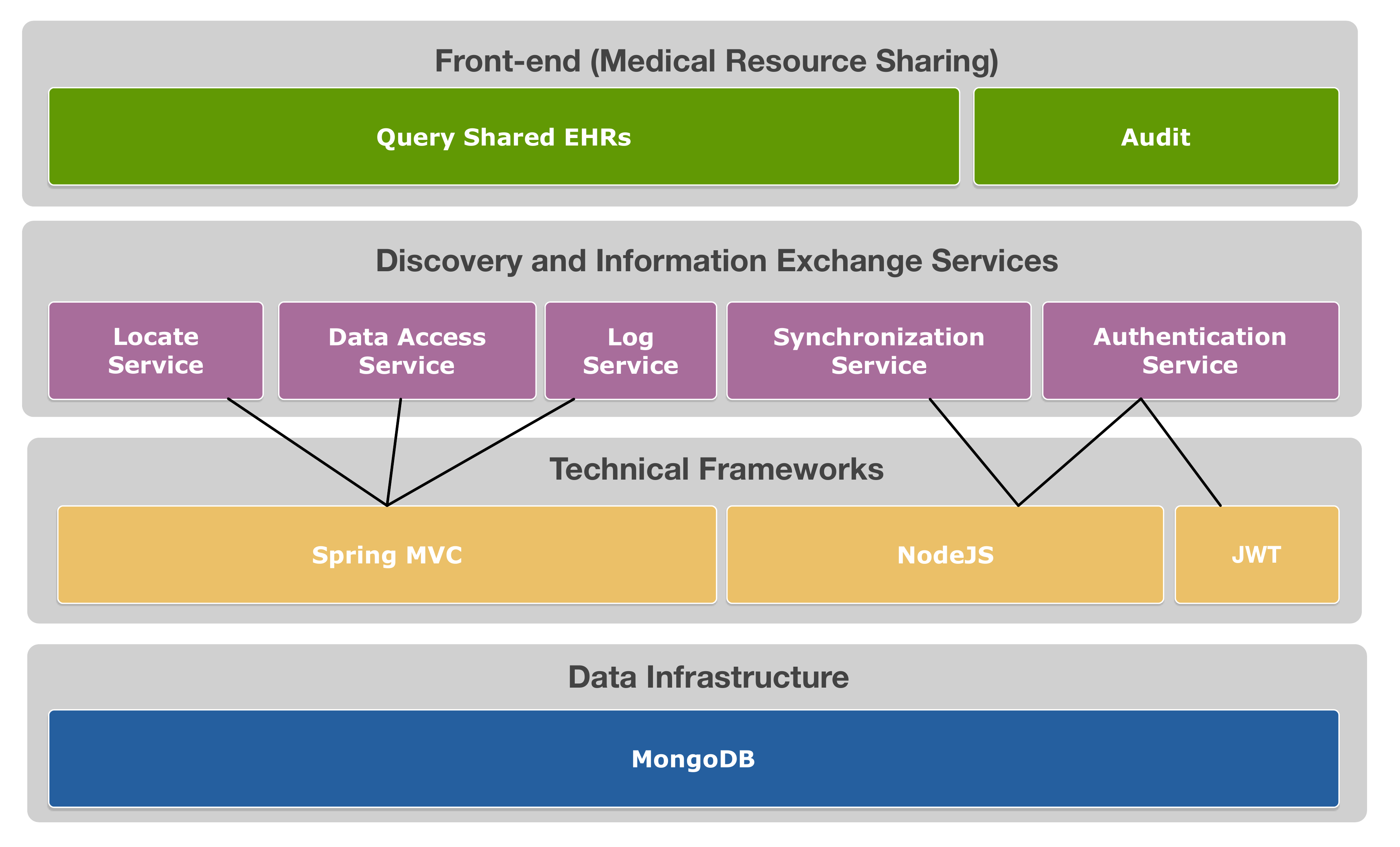}\\
\caption{Implementation Stack for the Medical Resource Sharing}
\label{techdetail}
\end{center}
\end{figure}
\noindent\textbf{Data Infrastructure Layer:} The data infrastructure, as shown in Fig.~\ref{techdetail}, shows the data storage based on MongoDB~\cite{Abramova2013} which is NoSQL database, more precisely a non-relational database. To deal with the complexity of medical data, it requires to have an adaptable format facilitating the data transformations easily across multiple sources. This approach overcomes the bottlenecks of traditional databases. Using MongoDB also helps mutability and scalability features of EHRs.

\vspace{.3cm}
\noindent
\textbf{Technical Framework Layer:} All the components described in our presented architectural models are implemented by the lightweight Java EE framework Spring~\cite{Gupta2010}.
The required two-way authentication service in the legacy EHR system is implemented as a RESTFul web service by NodeJS~\cite{Tilkov2010} and JSON Web Token (JWT)~\cite{Jones2014}. A RESTful service can be defined as a means to hold query parameters. Contrary to JavaEE, NodeJS has the advantage of utilizing low resources to support high concurrency, which is good at scaling it to industrial problems. While JWT is a compact, URL-safe approach for representing claims between two communicating ends. JWT provides foundational authentication service to RESTful web services.
Those two techniques guarantee the reliability and safety of the authentication process.

\vspace{.3cm}
\noindent
\textbf{Discovery and Information Exchange Services Layer:} 
This layer has three Spring MVC services and two web services for authentication and synchronization. The \emph{LocateService}, which is implemented using Spring MVC framework, identifies the required EHR location from the patients indexed in the MongoDB data infrastructure. The \emph{LocateService} locates the EHRs based on the search conditions and transmits it to the doctor. The \emph{DataAccess} is technically similar to \emph{LocateService} but functions differently. It retrieves relevant patient data from the identified source. The \emph{AuthenticationService} provides the authorization service to the patient when the doctor requests for a specific EHR. The authentication also requires a service that integrates legacy EHR system into the authentication process. The \emph{SynchronizationService} timely triggers the replication of the shared EHRs and updates the indexes in the patient indexing server. The \emph{LogService} provides the log and tracking services to avoid data breach and trace irregularities. The authentication component is deployed in all the hospitals. The EHR query component is deployed in all the hospitals to provide the data transmission service, and in the patient indexing server to provide the locating service. The \emph{SynchronizationService} is deployed in all the hospitals and the data center to replicate shared EHRs and update indexes. The \emph{LogService} is deployed on all servers because logs are generated and stored in the patient indexing server and all the other hospitals.

\vspace{.1cm}
\noindent
\textbf{Front-end Medical Resource Sharing Layer:} This layer combines all the described layers and directly utilizes the services available in the discovery and information exchange services layer. Through this front-end the end user poses a query to the shared EHRs resources and retrieves a list of resources against the targeted EHR. The \emph{Audit} service holds the system users accountable for their action in the system. More precisely, the doctor can distributively retrieve all the relevant records of the patient among all the hospitals while preserving patient privacy.

\subsection{Evaluation}
We deploy our prototype system in four different servers named HC, KW, UH, and PI (Patient Indexing).
The EHR data are retrieved from hemodialysis center of Kiang Wu hospital, and then also generated extensive testing data for other two hospitals. In our testing scenario, we have 10,000 hemodialysis data of 100 patients.
After accessing the shared data of these 100 patients with the different date range, \emph{MedShare} worked as desired. Furthermore, all shared data was successfully recorded by logging the data access requests and their providers.

\begin{figure}[!htb]
\begin{center}
\includegraphics[width=0.9\columnwidth]{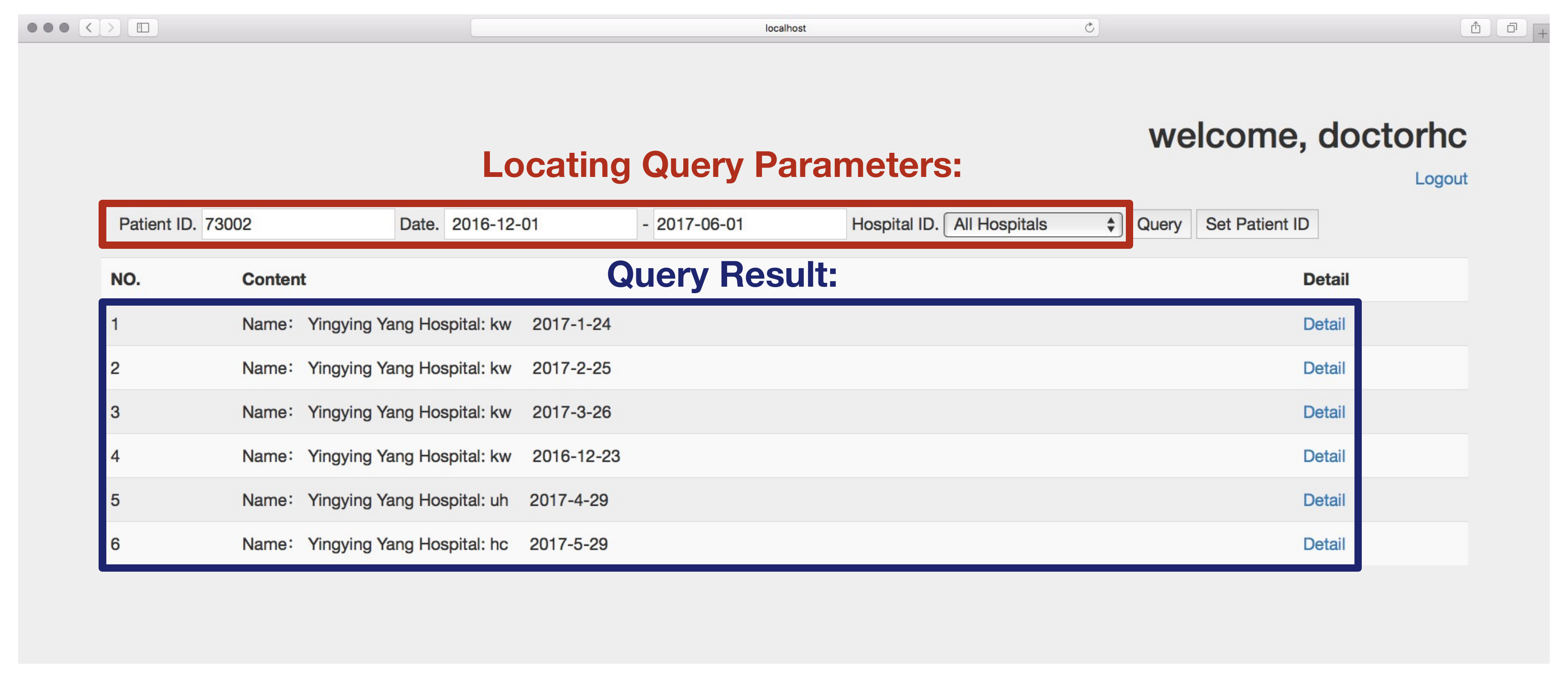}\\
\caption{A screenshot of the Query Execution Environment to Locate a Resource}
\label{LocatingService}
\end{center}
\end{figure}

Let us assume that a doctor in HC hospital requests all the hemodialysis records of a patient named Yang Yingying. This scenario is demonstrated by the Fig.~\ref{LocatingService} that shows a list of all the hemodialysis records of Yang Yingying that may further be individually viewed by the doctor by clicking the \emph{Details} link.

\begin{figure}[!htb]
\begin{center}
\includegraphics[width=0.9\columnwidth]{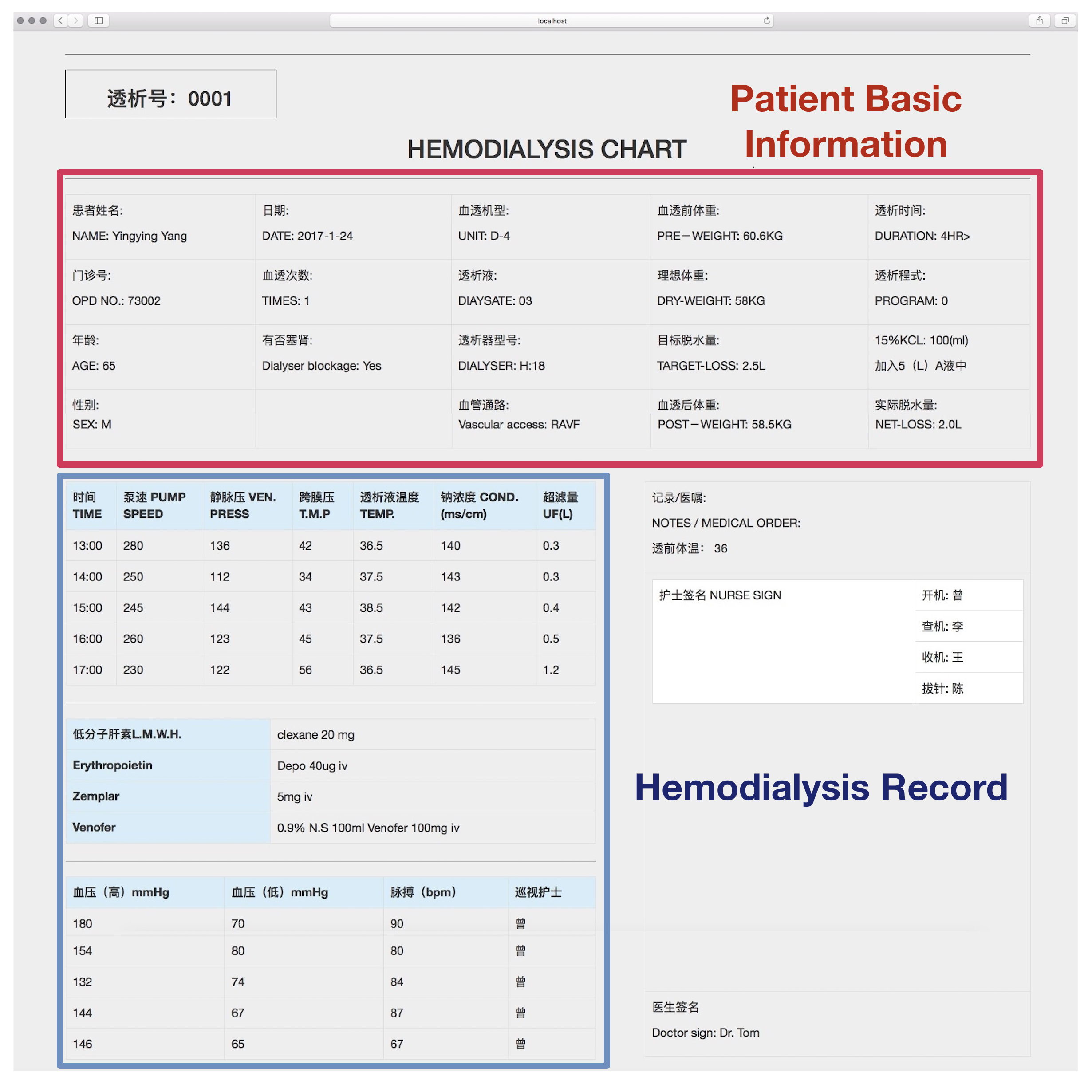}\\
\caption{A detailed Hemodialysis Report}
\label{EHRFormat}
\end{center}
\end{figure}

From the retrieved list of medical records, as given in Fig.~\ref{LocatingService}, a doctor can fetch the detailed record against any displayed link. For example, the EHR corresponding to Sep 30, 2015, as shown in Fig.~\ref{EHRFormat}. The detailed output includes two types of information: 1) the patient information, and 2) her hemodialysis record.

\begin{figure}[!htb]
\begin{center}
\includegraphics[width=0.9\columnwidth]{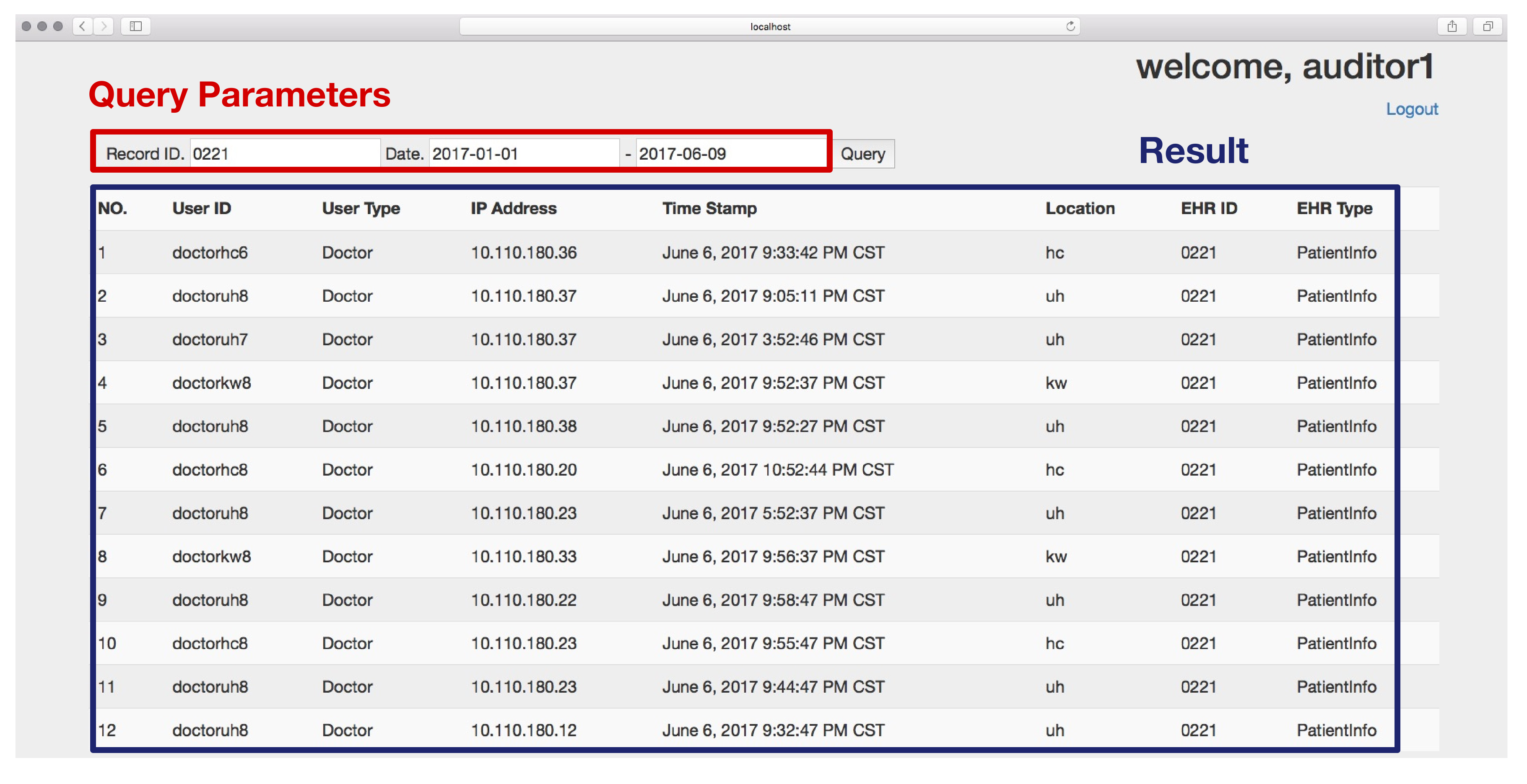}\\
\caption{Auditing access to a medical resource}
\label{audit}
\end{center}
\end{figure}

In order to be able to track the accessed data, \emph{MedShare} supports the administrator to track the logs and investigate the specific operations performed by the users on a patient record.
For example, when the tracker needs to know the accessing information about the EHR with ID 0221,
the tracking system is able to show all the relevant results between two dates, as demonstrated in Figure~\ref{audit}.
The prototype shows that the proposed architecture can deal with the sharing tasks of the hemodialysis EHRs among the Macau hospitals
without compromising the privacy requirements.

The distributed resource-sharing environment can be scaled up to include a significantly large number of medical health providers. However, in that case robust testing is proposed. Our resource sharing toolset lets its stakeholders to retain their data, which is otherwise a primary concern for a participating stakeholder. \emph{MedShare} provides a transparent platform by integrating legacy EHR systems that were developed different implementation techniques. To maintained the openness of the system, the participants chose their own interoperable data format through negotiations which may however be replaced with open standards such as HL7 and openEHR. From another technical perspective, we also need an in-depth analysis of data storage strategies.

\section{Conclusion and Future Perspectives}
\label{sec:Conclusion}

We presented a set of implementation guidelines for exchanging medical resources among autonomous healthcare providers. We negotiated a common data structure that sets forth the first step to allows for interoperability among the three disconnected hospitals in Macau. Applying standardized data formats, such as HL7, was a daunting task because of bilingual patient data storage in both English and Chinese. \emph{MedShare} ensured that participating healthcare providers have confidence in the system through their primary control over patient data. Our work endorses the fact that the exchange of medical information between independent hospitals is not only limited to technical issues but economic and political issues are equally important.

Our experience with developing interoperable systems advocates a gradual replacement of legacy EHRs systems. \emph{MedShare} preserves patient privacy by two-way authentication process that collects patient consent before any data authorization is made. To integrate patient consent into a data-sharing scenario, our system takes the advantage of national identification cards that are swiped by the patients during their medical visits. All patients in a hospital are uniquely identified by their identity numbers, which are hashed in the data indexing process. The patient indexing technique enables a more secure data exchange environment and develops a sense of safe cooperation between hospitals.

Our future work includes developing an intense auditing process over shared medical data. To this end, we also aim to study potential attacks on the deployed system. In data sharing scenarios where multiple languages are used to store, process and communicate data, language-dependent and unified data formats are not directly applicable. This necessitates an additional work to tackle interoperable systems using two or more languages. Thereby, both syntax and semantics play an important role to develop such system. We aim to increase the number of hospitals in our interoperable resource sharing network. We also plan to report our findings based on the scalability and openness of the system. Robust evaluation studies are needed to evaluate non-functional aspects of such systems including scalability, heterogeneity, resource management, transparency, openness and performance analysis.


\section*{Acknowledgment}
This work was supported by the Macau Science and Technology Development Fund (FDCT) (No. 103/2015/A3 and 018/2011/A1) and the National Natural Science Foundation of China (NSFC) (No. 61562011 and 61672435).




\bibliographystyle{plain}
\bibliography{IEEEabrv}

\end{document}